# Challenges and Opportunities of SRF Theory for Next Generation Particle Accelerators

Alex Gurevich

Department of Physics and Center for Accelerator Science, Old Dominion University, Norfolk VA 23529, USA.

Takayuki Kubo

High Energy Accelerator Research Organization (KEK), Tsukuba, Ibaraki 305-0801, Japan

James A. Sauls

Northwestern University, Department of Physics and Astronomy, Evanston, IL 60208, USA

# Executive Summary

Recent technological advances have produced superconducting niobium cavities for which the maximum accelerating field gradient reached about 52 MV/m at 2 K. This implies that the peak magnetic field at the equatorial cavity surface is close to the dc superheating field $H_s \approx$ 240 mT and the best niobium cavities are close to the dc pairbreaking field limit. At the same time, materials treatments such as low-temperature baking and nitrogen infusion have boosted the quality factors to $Q = (2-4) \times 10^{10}$ in the field range of $0 < H < 150$ mT. These achievements pose a question whether the best Nb cavities are already close to the fundamental limit of SRF performance and reaching the accelerating gradients ~ 100 MV/m at 2K or ~ 50 MV/m at 4.2K may require new SRF materials (like Nb$_3$Sn) with higher superheating magnetic fields than Nb. In this white paper we show that currently this question cannot be answered since the theoretical SRF performance limits at GHz frequencies are unknown and their current estimates are primarily based on unsubstantiated extrapolations of classical old results: 1. Mattis-Bardeen theory of surface resistance which is only applicable at low fields $H \ll H_s$, 2. the theory of the dc superheating field at temperatures close to the critical transition temperature $T_c$ and thus hardly applicable at GHz frequencies and low temperatures $T \ll T_c$ relevant to the SRF cavities.

We suggest a synergetic program to establish the theoretical SRF performance limits at GHz frequencies using modern theories of nonequilibrium superconductivity under strong electromagnetic field which have been developed in the last 20 years. These theories will be used to calculate the main parameter of merit of SRF cavities: the quality factor $Q(H,T,\omega)$ and its dependencies on the field amplitude, temperature and frequency, which would allow us to understand how far the SRF cavity performance could be pushed from the current state-of-the-art. Given that $Q(H,T,\omega)$ is determined by multiple mechanisms operating on very different length scales, we'll address the interconnected problems of a nonlinear surface resistance, rf losses of vortices trapped in the cavity, the effect of materials defects and surface topography, and opportunities to boost the SRF performance by surface nano-structuring, impurity management and multilayers. We suggest the following key directions of theoretical SRF research to address the HEP goals of boosting the performance of the next generation particle accelerators:

- Establishing the Q limit: mechanisms of nonlinear surface resistance and the residual resistance in a nonequilibrium superconductor under a strong RF field.
- Establishing the SRF breakdown field limit: Dynamic superheating field and its dependencies on frequency, temperature and concentration of impurities.
- RF losses due to trapped vortices and extreme dynamics of ultrafast vortices driven by strong rf Meissner currents in SRF cavities.
- Optimization of SRF performance due to surface nano-structuring of the cavity surface: SIS multilayers and impurity management.

These theoretical issues have been brought to focus only recently due to the advances in the SRF cavity technology and quantum information technologies involving transmon superconducting circuits. Addressing these issues will require coordinated efforts and collaboration of different theoretical groups working on SRF problems at Old Dominion University, Northwestern University, KEK, Cornell, ANL and others. Besides the establishing the fundamental limits of the accelerating gradient and the quality factors, this work will develop numerical codes to calculate the field dependence of $Q(H_a)$. These theoretical activities will be performed in collaboration with SRF experimental groups to meet the needs of the HEP community. This program can provide a platform for training graduate students and young scientists who will be working at the forefront of theoretical research while collaborating with the leading SRF experimental and technological groups at National Laboratories and Universities. Implementation of this research will require extensive numerical simulations some of which may require the access to the computational resources of National Laboratories.

# Introduction

Superconducting radio frequency (SRF) technology based on Nb resonant cavities has been instrumental for many high-energy particle accelerators. Over the past decades, the RF performance of bulk Nb cavities has continuously improved with material and surface treatments, such as, low-temperature baking [1-2] and nitrogen infusion and other materials treatments [3-7]. As a result, the current state of the art Nb cavities can have the quality factors $Q = (2-6) \times 10^{10}$ @ 2K and 1-2 GHz and produce the accelerating gradients up to 52-54 MV/m at which the peak magnetic field at the cavity equator reaches 200-240 mT close to the dc superheating field of Nb. Such exemplary SRF performance poses the question whether the Nb cavity technology has approached a fundamental limit determined by the physics of superconductivity or further significant developments are still possible. It turns out that an unambiguous answer to this question cannot be given because the superconductivity limits under strong RF field are still unknown and the conventional evaluations of these limits by extrapolating the well-known dc field limits and the low-field Mattis-Bardeen surface resistance to strong RF fields are not justified. This problem brings about the necessity of in-depth theoretical investigations of fundamental RF performance limits of superconductors under strong RF electromagnetic fields relevant to particle accelerators. Such investigations should address many outstanding issues of superconductivity theory, particularly the dynamic superheating field which controls the field limit of superconductivity breakdown, nonlinear surface resistance and dissipation of trapped vortices in strong RF field which control the maximum Q factors and the physics of residual resistance which control the limit of Q at low fields. Solving these challenging problems would not only establish the much-needed fundamental SRF limits but also help guiding the materials optimization to boost the cavity performance and extend the high energy physics research to other areas, particularly the emerging quantum information technologies involving high-Q SRF cavities.

## Background

The theoretical background for the SRF cavity accelerator technology was established by the Bardeen-Cooper-Schrieffer (BCS) theory of superconductivity which explained both the Meissner effect and the extremely low power loss in superconductors under weak RF electromagnetic fields at temperatures $T$ much lower than the critical temperature $T_c$. Particularly, the Mattis-Bardeen and Abrikosov-Gor'kov theories gave a general nonlocal between the current density and the applied weak magnetic vector potential and showed the important effect of atomic impurities [9-12]. In an impure superconductor with the mean free path $l$ shorter than the coherence length $\xi_0$ which quantifies a size of the Cooper pair, the SRF electromagnetic response can be described by a frequency-dependent surface impedance $X(\omega) = R_s(\omega) - iX_s(\omega)$. Here the reactive part $X_s = 1/\omega\mu_0\lambda$ describes the Meissner effect, and the surface resistance $R_s$ at low temperatures and frequencies ($T \ll T_c$, $\hbar\omega \ll \Delta$) relevant to the SRF cavities is given by [13,14]:

$$R_s = \frac{\mu_0^2 \omega^2 \lambda^3 \Delta}{\rho_n k_B T} \ln\left[\frac{9k_B T}{4\hbar\omega}\right] e^{-\Delta/k_B T}, \qquad (1)$$

where $\Delta = 1.8 k_B T_c$ is the superconducting gap, $\lambda$ is the magnetic London penetration depth, and $\rho_n$ is a resistivity in the normal state. The surface resistance determines the quality factor $Q = CR_0/R_s$, the main parameter of merit for SRF cavities, where $R_0 = 377$ ohm is the vacuum impedance and $C \sim 1$ is a geometric factor. For the best Nb cavities at 2 K and 1-2 GHz, the BCS surface resistance (1) is can be as low as $10-20$ nohm, which translates to $Q \sim (2-6) \times 10^{10}$. Equation (1) and its full Mattis-Bardeen generalization for arbitrary $T, \omega$ and $l$ describe the observed dependencies of $Q(T,\omega,l)$ on frequency, temperature, and the mean free path. It has often been used to analyze Q at high RF fields at which Eq. (1) is not applicable. Moreover, Eq. (1) does not explain a temperature-independent residual resistance $R_i$ which has been observed below 1.5-1.7 K, so it can hardly provide a solid background for the evaluation of Q limits, particularly at high RF fields.

The key advantages of SRF cavities are fully revealed only if they operate in the vortex-free Meissner state so the field limit is determined by the superheating field $H_s(T)$ – the maximum field which the Meissner state can withstand before it becomes unstable with respect to penetration of vortices. The dc superheating field at $T \approx T_c$ has been calculated by many groups by numerically solving the Ginzburg-Landau (GL) equations [15-17], but $H_s(T)$ at the most relevant for the SRF cavities low temperatures $T \ll T_c$ remains largely unknown except for the limit of a large GL parameter $\kappa = \lambda/\xi \gg 1$ [18-22]. Yet the results for $H_s(T)$ at $\kappa \gg 1$ cannot be used to evaluate the dc field limits for Nb which has $\kappa \simeq 1$, so the calculations of $H_s(T)$ at all temperatures and GL parameters are needed. Moreover, it is unclear how far the dynamic superheating field $H_d(T)$ at GHz frequencies can be from the static $H_s(T)$. Calculations of the dynamic superheating field are much more complicated as compared to the static $H_s$, as they require solving a full set of dynamic BCS equations which describe intertwined effect of RF current pairbreaking and nonequilibrium kinetics of quasiparticles under a strong RF field. The first results in that direction are promising, showing that the dynamic superheating field near $T_c$ is $\sqrt{2} \approx 1.41$ times larger than the static $H_s$ at GHz frequencies for which $\omega\tau > 1$, where $\tau(T)$ is the energy relaxation time of electrons on phonons [23]. However, the key question about the dynamic superheating field at $T \ll T_c$ remains open.

One of the serious theoretical challenges is to develop a theory of $Q(T,H_a)$ at strong RF fields $H(t) = H_a \cos(\omega t)$ which drive a superconductor in a nonequilibrium state. As a result, a theory of nonlinear electromagnetic response of superconductors under strong RF fields becomes far more complex than the Mattis-Bardeen theory which is only applicable at very weak RF fields. The theory includes a full set of coupled dynamic equations for the superfluid density, superconducting gap, spectral functions, and kinetic equations for the quasiparticle distribution function to account for different channels of energy and momentum relaxation in electron-phonon or electron-electron inelastic collisions [12,24-26]. It was shown [13,27] that some of these mechanisms such as the effect of current broadening of the quasiparticle density of states [28-34] can contribute to the negative $Q(H_a)$ slope observed on alloyed Nb cavities [3-8]. Other contributions can result from atomic two-level systems caused by the interstitial hydrogen [35] and trapped vortices [36]. A theory of nonlinear surface resistance which incorporates these issues should address many outstanding issues of nonequilibrium superconductivity and eventually develop efficient numerical codes to calculate the field dependence of $Q(H_a)$.

One of the unresolved theoretical issues is the physics of a residual surface resistance $R_i$ which is not described by Eq. (1). Addressing the problem of nonzero $R_i$ requires going beyond the conventional Mattis-Bardeen and Abrikosov-Gorkov theories of weak electron scattering on sparse noncorrelated impurities and develop a theory of surface resistance, taking into account multiple scattering of electrons on spatially correlated impurities, as well as dynamic inelastic effects in scattering on impurities. The problem of residual resistance is closely related to a finite density of quasiparticle states at energies below Δ (the so-called subgap states) which have been often observed by scanning tunneling spectroscopy (STM) [37,38]. Addressing the mechanisms of the residual resistance is important not only for establishing the upper limit of quality factors in SRF cavities but also for quantum information technologies involving superconducting circuits in which $R_i$ determines the lower dissipation limit and decoherence times in qubits at very low temperatures.

Another significant contribution to RF losses can come from vortices trapped in the cavity during its cooldown through $T_c$ (see, e.g., Ref. 39 and references therein). Trapped vortices can increase the residual surface resistance so theoretical investigations of losses caused by vibrating elastic vortex lines driven by surface RF currents and pinned by materials defects are necessary. Both theoretical and numerical calculations of vortex RF losses [40-43] can provide a guidance for minimization of low-field RF losses due to trapped flux by optimizing pinning defect nanostructure and concentration of impurities at the cavity surface. Hotspots caused by trapped vortices at the cavity surface can ignite premature superconductivity breakdown below the superheating field. It is therefore important to develop a theory of RF loses of ultrafast vortices driven by very strong RF currents for which pinning becomes ineffective [43] and evaluate an acceptable density of trapped vortices which do not significantly deteriorate the SRF performance.

The issues outlined above suggest the following key directions of theoretical SRF research to address the HEP goals of boosting the performance of the next generation particle accelerators:

- Establishing the Q limit: mechanisms of nonlinear surface resistance and the residual resistance in a nonequilibrium superconductor under a strong RF field.
- Establishing the SRF breakdown field limit: Dynamic superheating field and its dependencies on frequency, temperature and concentration of impurities.
- RF losses due to trapped vortices. Extreme dynamics of ultrafast vortices driven by strong rf Meissner currents characteristic of the SRF cavities.
- Optimization of SRF performance due to surface nano-structuring of the cavity surface: SIS multilayers and impurity management.

This program includes many outstanding theoretical issues which have been brought to focus only recently due to the advances in the Nb cavity technology and the progress in quantum information technologies involving transmon superconducting circuits. This challenging program will require coordinated efforts and collaboration of different theoretical groups working on SRF problems at Old Dominion University, Northwestern University, KEK, Cornell, ANL and others. In addition to the establishing of the fundamental limits of accelerating gradients and quality factors of SRF cavities, the outcome of this work will be the development of efficient numerical codes to calculate the field dependence of $Q(H_a)$. This theoretical work will be performed in collaboration with SRF experimental groups at FNAL, JLab, SLAC, Europe, Japan and China to guide the ongoing optimization of SRF cavities and get feedback from experimental results to help refine the theory and meet the needs of the HEP community.

## Dynamic superheating field and the limit of accelerating gradient

The static superheating field $H_s(T,\kappa)$ at $T \approx T_c$ has been calculated by many authors by solving the Ginzburg-Landau (GL) equations [15-17]. These results, which have been widely used to evaluate the maximum breakdown fields of SRF cavities at $T \ll T_c$, show that $H_s(T,\kappa)$ decreases monotonically with the GL parameter $\kappa = \lambda/\xi$, reaching $H_s \approx 1.2 H_c$ at $\kappa \approx 1$ (clean Nb) and approaching $H_s \approx 0.745 H_c$ at $\kappa \gg 1$ (Nb$_3$Sn, dirty Nb, pnictides). Here the thermodynamic magnetic field $H_c = \phi_0/2^{3/2}\pi\mu_0\lambda\xi$ is nearly independent of the concentration of nonmagnetic impurities [44]. At $H = H_s$ the Meissner state becomes unstable with respect to infinitesimal *inhomogeneous* perturbations of current and the order parameter with the wavelength $\sim \xi^{3/4}\lambda^{1/4}$ along the surface and decaying over the length $\sim \sqrt{\lambda\xi}$ perpendicular to the surface [15-17]. This feature represents the main challenge in the calculations of $H_s(T,\kappa)$ by solving the quasi-classical Eilenberger or Usadel equations [12] at low temperatures $T \ll T_c$ of interest to the SRF cavities. So far $H_s(0)$ has only been calculated in the limit of $\kappa \to \infty$ in which $H_s = 0.84\, H_c$ in the clean limit $l \gg \xi$ [18,19] and for arbitrary concentrations of nonmagnetic and magnetic impurities [20] giving $H_s(0) = 0.795 H_c$ in the dirty limit $l \ll \xi$ [22]. Moreover, $H_s$ can be significantly reduced by pairbreaking magnetic impurities [20] of finite quasiparticle lifetime quantified by the Dynes parameter $\Gamma$ [22]. As a result, regions with enhanced concentration of magnetic impurities or the parameter $\Gamma$ at the surface can locally degrade $H_s$ causing local penetration of vortices and hotspots triggering thermal quench of the cavity. Yet, the establishing of the dc field limits for SRF cavities at $T \ll T_c$ require full microscopic calculations of $H_s(T, \kappa, l)$ which has not yet been done.

Establishing the rf field limits of SRF cavities is a very challenging theoretical problem of stability analysis of a superconductor in a nonequilibrium state under a strong RF field for which the dynamic superheating field $H_d(T, \omega, \kappa, l)$ becomes a function of the rf frequency. Calculation of $H_d(T, \omega, \kappa, l)$ then requires numerical solution of very complicated coupled nonlinear equations for the superconducting order parameter and a kinetic equation for a distribution function of nonequilibrium quasiparticles, taking into account elastic scattering on impurities and inelastic scattering on phonons and electrons which provide the transfer of rf power absorbed by quasiparticles to the crystal lattice and then to the He bath [12]. For instance, $H_d(T, \omega, \kappa, l)$ becomes dependent on the relation between the rf period and the relaxation time of hot quasiparticles on phonons, $\tau_E \sim (\hbar/k_B T)(c_s/v_F)^2 (T_F/T)^2$, where $c_s$ is the speed of sound, and $v_F$ and $T_F = mv_F^2/2k_B$ are the Fermi velocity and temperature, respectively [12]. Because $\tau_E \propto T^{-3}$ increases rapidly as the temperature decreases, the rf field drive quasiparticles out of equilibrium at low $T$ for which $\omega\tau_E(T) > 1$ at GHz frequencies of SRF cavities. The first calculation of $H_d(T, \omega, \kappa, l)$ at $T \approx T_c$ [23] showed that at $\omega\tau_E(T) \gg 1$ the dynamic superheating field $H_d = \sqrt{2}H_s$ can exceed the static $H_s$ by 41 % but the relation between $H_d$ and $H_s$ at $T \ll T_c$ depends essentially on the mean free path. In a dirty superconductor with $l < \xi$ the density of nonequilibrium quasiparticles at $H = H_s$ remains exponentially small [20,30] so they do not really affect the breakdown of superconductivity by rf field with $\hbar\omega \ll \Delta$ (in Nb cavities $\hbar\omega/\Delta \sim 10^{-2}$). As a result, $H_d(T)$ at $T \ll T_c$ becomes close to $H_s(T)$ even if $\omega\tau_E(T) \gg 1$ [23]. By contrast, in a clean superconductor with $l \gg \xi$, the quasiparticle gap vanishes before the field amplitude reaches $H_s$ [20,30] so the density of nonequilibrium quasiparticles is not small and they can likely increase $H_d$ relative to $H_s$. The challenging calculation of the dynamic superheating field in a clean superconductor at $T \ll T_c$ based on the well-established theoretical framework of dynamic equations for Green's functions and kinetic equations for a nonequilibrium superconductor [12,24-26] would therefore be a very important step to establish a true SRF field limit.

## Nonlinear surface resistance and the limits of the quality factor $Q(H_a)$.

The field-dependent quality factor $Q(H_a) = CR_0/R_s(H_a)$, the main parameter of merit of SRF cavities, is determined by the nonlinear surface resistance $R_s(H_a)$. The microscopic calculation of $R_s(H_a)$ at high rf field and low temperatures involves the account of multitude of interconnected effects including electron and phonon overheating and the rf current pairbreaking which causes temporal oscillations of the quasiparticle density of states $N(\epsilon, t)$ and the superconducting pair potential $\Delta(t)$ affecting the kinetics of nonequilibrium quasiparticles which transfer the absorbed rf power to phonons. Solving the equations of nonequilibrium superconductivity [12,24-26] to calculate the nonlinear surface resistance is even more challenging than that the problem of the dynamic superheating field discussed above, as $R_s(T, H_a)$ depends crucially on fine details of the nonequilibrium distribution function of quasiparticles in the subgap energy region $|\epsilon| < \Delta$. While self-consistent calculation of $R_s(T, H_a)$ has not yet been done, theoretical models [14,24] have revealed some of the essential mechanisms of the field dependence of $R_s(H_a)$ which may account for the extended $Q(H_a)$ rise observed on Nb cavities alloyed with N, O or Ti [3-7]. One of such mechanisms results from the well-known effect of broadening of the gap peaks in the density of states (DOS) $N(\epsilon, t)$ at $\epsilon \approx \Delta$ by current, which was predicted theoretically long ago [28-33] and observed directly by STM [34]. It was shown that the DOS broadening can describe the observed negative Q slope in Nb cavities [14,24], although a complete theory of the full dependencies of $R_s(H_a, T, \omega, l)$ on the field amplitude, temperature, rf frequency and the mean free path has not yet been developed. A well-established theoretical framework for such calculations of nonlinear electromagnetic response of a dirty superconductor is based on the time-dependent Usadel equations [12,24-26] which account for the effects of rf current pairbreaking on spectral functions and $\Delta(t)$ coupled with kinetic equations for nonequilibrium quasiparticles.

Besides the field dependence of $R_s(H_a, T, \omega, l)$ resulting from the BCS quasiparticle nonlinearities in an ideal superconductor, significant contributions to the $Q(H_a)$ slope can also come from the materials defects. For instance, it was shown that atomic two-level systems associated with interstitial hydrogen impurities could account for the initial $Q(H_a)$ rise at low fields [35]. Another possible contribution to the field dependence of $R_s(H_a)$ and the $Q(H_a)$ rise can come from large-amplitude vibrations of vortices trapped by materials defects [36]. A generic contribution to the field dependence of $Q(H_a)$ is due to the dependence of the London penetration depth on the field amplitude (a so-called nonlinear Meissner effect caused by the reduction of $\Delta(t)$ by rf current [26]). This effect is amplified by weak-linked grain boundaries which are particularly relevant in polycrystalline $Nb_3Sn$ [45]. Theoretical descriptions of these mechanisms would allow us to understand the field dependence of the quality factor $Q(H_a)$ and its limit at intermediate fields at which $Q(H_a)$ can significantly exceed the low-field value given by Eq. 1 of the Mattis-Bardeen theory.

**Residual surface resistance, impurities, grain boundaries and trapped vortices**

A temperature-independent residual surface $R_i$ resistance which has been often observed on superconducting materials, including SRF cavities [1,2] cannot be explained by the Mattis-Bardeen theory in which $R_i = 0$. This is because the density of states in the BCS model vanishes at energies below the gap: $N(\epsilon) = 0$, $\epsilon < \Delta$. The residual resistance appears if there are subgap quasiparticle states at energies $\epsilon < \Delta$. Such subgap states which have observed in numerous STM experiments [37] are described in the literature by the Dynes phenomenological model in which $N(\epsilon) = N_0 Re\left[(\epsilon + i\Gamma)/\sqrt{(\epsilon + i\Gamma)^2 - \Delta^2}\right]$, where $N_0$ is the DOS in the normal state and the parameter $\Gamma$ quantifies the quasiparticle lifetime $\hbar/\Gamma$. The residual resistance is then [14,46]:

$$R_i = \frac{\mu_0^2 \omega^2 \lambda^3 \Gamma^2}{2\rho_n(\Delta^2 + \Gamma^2)} \qquad (2)$$

where $\rho_n$ is the normal state resistivity. For $\lambda = 40$ nm and $\rho_n = 1$ nohm*m, Eq. 2 gives $R_i = 10$ nohm in Nb at 1.5 GHz at $\Gamma/\Delta = 0.05$. Different contributions to $\Gamma$ have been suggested in the literature (see, e.g., a discussion in Ref. 46), but none of them has been unambiguously identified as a prime source of residual resistance in Nb cavities. One of the challenges of SRF theory is to generalize the conventional Abrikosov-Gorkov theory of weak uncorrelated impurity scattering [10,11] to go beyond the Born approximation and include effects of spatial correlations of impurities which could produce the subgap states [47]. A contribution to the residual resistance can also come from scattering of electrons on weak-linked grain boundaries. This effect can be essential in polycrystalline $Nb_3Sn$, but can hardly in Nb given that significant values of $R_i \simeq 10$ nohm have been observed on large-grain Nb cavities at 1.5 GHz and 2 K.

A well-documented contribution to the residual resistance comes from vortices trapped in the cavity during its cooldown through $T_c$. The rf losses produced by sparse trapped vortices have been calculated in Refs. 41-43. For instance, extensive numerical simulations of a curvilinear elastic vortex driven by the surface rf current and pinned by uncorrelated pinning centers have been performed to calculate the vortex losses and the residual resistance [43]. It turns out that the vortex losses at weak rf fields $H_a \ll H_c$ can be reduced by optimizing the pinning potential and the impurity mean free path at the surface. Yet this optimization becomes ineffective at high fields $H_a \gtrsim 10$ mT which induce of 5% of the depairing current density $J_d = H_s/\lambda \simeq 4 \times 10^{12}$ A/m$^2$ at the surface, orders of magnitude higher than typical depinning current densities $J_c \sim 10^8$ A/m$^2$ in Nb [39,48]. Moreover, large densities of pinning centers can significantly increase rf losses and cause current blocking effects.

**Hotspots and extreme dynamics and losses of trapped vortices**

Trapped vortices cluster around pinning defects and produce hotspots at the cavity surface which have been revealed by temperature mapping [1,2,41]. In cavities the tips of these vortices are exposed to very strong rf currents with densities $J = H_a/\lambda$ much higher than any realistic depinning current density $J_c$. Scanning SQUID microscopy has shown that vortices driven by such strong currents can move as fast as $10 - 20$ km/s, much faster than either the speed of sound and the maximum superfluid velocity $\Delta/p_F$ of the supercurrent [49]. According to numerical simulations of the time-dependent Ginzburg-Landau (TDGL) equations, the normal vortex core at such high velocities extends along the direction of motion and the viscous drag coefficient $\eta(v)$ becomes dependent on the vortex velocity $v$ [49]. This issue was addressed by Larkin and Ovchinnikov (LO) who showed that the drag coefficient $\eta(v) = v_0/(1 + v^2/v_0^2)$ decreases with $v$, resulting in jumps on the V-I characteristics [24]. Here $\eta_0 = \phi_0^2/2\pi\xi^2\rho_n$ is the Bardeen-Stephen drag coefficient at small $v$ [44], and the critical LO $v_0(T)$ decreases as $T$ decreases and is typically of the order of $0.1 - 1$ km/s [36,43]. The extreme dynamics of fast Josephson vortices driven by strong current along a grain boundary in a film coating and a dynamic transition of the vortex to a phase slip was considered in Ref. 50.

Dissipation of a vortex perpendicular to the cavity surface was calculated for different spatial distributions and strengths of pinning centers for both the Bardeen-Stephen and the velocity-dependent LO drag coefficient [43]. The residual surface resistance $R_i(H_a)$ obtained by averaging over different pin configurations increases smoothly with the field amplitude at small $H_a$ and levels off at higher fields consistent with a nearly linear increase of $R_i(H_a)$ which has been observed on some Nb cavities [51]. For strong pinning, the LO decrease of $\eta(v)$ with $v$ can result in a nonmonotonic field dependence of $R_i(H_a)$ which decreases with $H_a$ at higher fields but cause a runaway instability of the vortex for weak pinning. Overheating of a moving vortex can also produce the LO-like velocity dependence of $\eta(v)$ but mask the decrease of the surface resistance with $H_a$ at a higher density of trapped vortices [43]. Here the SRF cavities provide a unique experimental tool to study the rich but poorly understood physics of fast vortices driven by strong Meissner currents. The power dissipated by moving vortices is determined by a multitude of complex nonequilibrium effects caused by quasiparticles trapped in the stretched vortex core, which represent yet another challenge for the SRF theory.

**Materials optimization: Surface nano-structuring, multilayers and impurity management**

Based on theoretical understanding of the complex physics behind the exemplary performance of accelerator cavities, one can suggest new ways to further increase the SRF performance limits by modifying the material at the inner cavity surface. These suggestions include the formation of a dirty layer with higher concentration of nonmagnetic impurities at the surface [21], modification of properties of a proximity-coupled metallic suboxide NbO layer covering the Nb surface [46], formation of a layer with dilute magnetic impurities [46] or multilayer nanostructuring in which thin layers of superconductors (S) with $H_c$ higher than $H_c^{Nb}$ of Nb are separated by thin dielectric (I) layers [56-62]. The latter allows one to significantly increase the field onset of penetration of vortices and use superconductors with $T_c$ and $\Delta$ larger than in Nb but have the lower critical field smaller than $H_{c1}^{Nb}$, particularly Nb$_3$Sn or superconducting pnictides.

As was shown in Refs. 52-55, the surface resistance $R_s$ can be reduced by engineering an optimum broadening of DOS peaks using various pair-breaking mechanisms which decrease $T_c$ and $H_c$ in the bulk. For example, a layer of dilute magnetic impurities spaced by a few $\mu$m can decrease the surface resistance by as much as 40% below the Mattis-Bardeen $R_s$ for an ideal surface [46]. A similar reduction of $R_s$ can also be achieved by tuning the Dynes $\Gamma$ parameter or the thickness and conductivity of the metallic suboxide layer and its contact resistance by different heat treatments of the material [52-55]. The pair-breaking mechanisms caused by the normal suboxide, finite Dynes parameter or magnetic impurities can strongly modify the field-dependent nonlinear $R_s(H_a)$. For example, a proximity-coupled normal layer at the surface can shift the minimum in $R_s(H_0)$ to either low and high fields and can reduce $R_s$ below that of an ideal surface. At the same time, the, proximity coupled metallic suboxide layer increases the residual surface resistance [46]. These results may explain why quality factor $Q(H_0)$ can be so sensitive to the materials processing, such as low-temperature baking [1,2].

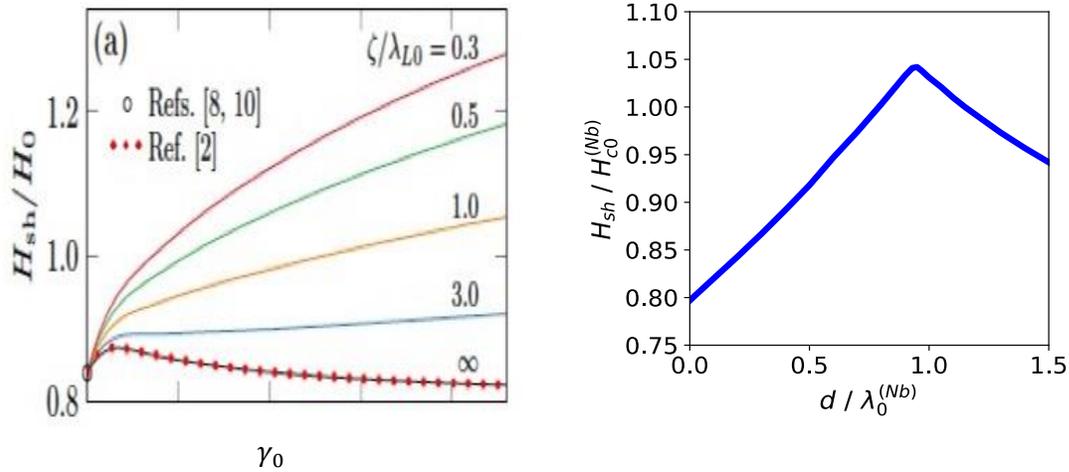

Fig. 1. Left: Enhancement of the superheating field by a dirty layer at the surface of different thickness $\zeta$ for the exponential profile of impurity concentration [21]. Right: Increase of $H_s$ by a uniform dirty Nb layer of thickness $d$ on the surface of the Nb cavity [62].

Shown in the left panel of Fig. 1 is an example of increasing $H_s$ by a thin dirty layer with enhanced impurity scattering rate modeled by the exponential profile $\gamma(x) = \gamma_0 \exp(-x/\zeta)$, where $\zeta$ is the thickness of the layer and $\gamma_0$ is the magnitude of the extra scattering [21]. Here a thin dirty layer with large $\gamma_0$ can increase $H_s$ by as much as 40% relative to the case of uniform distribution of impurities in the dirty limit [20]. The right panel shows a similar enhancement of $H_s$ by a uniform dirty layer of thickness $d$ for which the maximum $H_c$ is reached at the optimum thickness $d \approx 0.9\lambda$ due to a current counterflow induced by the cleaner Nb substrate in a dirty layer [57,58].

For a superconducting multilayer, there is an optimal multilayer thickness $d = d_c$ at which it screens the maximum field $H_m$ [58]:

$$H_m = \left[H_s^2 + \left(1 - \frac{\lambda_0^2}{\lambda^2}\right) H_{s0}^2\right]^{1/2}, \qquad d_c = \lambda \ln\left[\mu + \sqrt{\mu^2 + k}\right]. \qquad (3)$$

Here the S layers with a bulk superheating field $H_s$ and the penetration depth $\lambda$ is deposited onto the Nb cavity with superheating field $H_{s0}$ and the penetration depth $\lambda_0$, where $\mu = \lambda H_s/(\lambda + \lambda_0)H_{s0}$, and k = (λ - $λ_0$)/(λ + $λ_0$). The optimized multilayer with d = $d_c$ can screen the field $H_m$ exceeding the bulk superheating fields of both Nb and S-layers if λ of the S-layer is larger than $λ_0 ≈ 40$ nm of clean Nb. This enhancement of $H_m$ results from counterflow induced in the S layer by the Nb substrate if λ > $λ_0 ≈ 40$ nm. Similar conclusions were also inferred from numerical simulation of the Ginzburg-Landau at T close to $T_c$ [62]. The theory shows that a current counterflow induced by the bulk Nb reduces the RF losses at high fields, mitigating the high-field-Q-slope. For a dirty surface layer, the increase of the penetration depth with rf field (the nonlinear Meissner effect [26]) becomes essential, as a thicker layer becomes necessary to protect the bulk Nb as compared to the London theory. The calculation shows [62] that this increases the optimum thickness by ∼ 10% as compared to Eq. 3. Addressing the vortex losses in multilayers under strong RF fields requires consideration of penetration of small vortex semi-loops at the surface defects and blocking these semi-loops by dielectric layers [58,60]. This dynamic nonlinear process could be calculated by numerical simulations of the time-dependent Ginzburg-Landau equations, which would give an opportunity to evaluate the effectiveness of the multilayer coating to block the detrimental thermomagnetic avalanches resulting from penetration of vortices from surface defects at $H_a > H_{c1}$. Another important issue is a theoretical analysis of the effect of weak-linked grain boundaries on the SRF performance, which can be particularly relevant for $Nb_3Sn/Al_2O_3$ multilayers [63,64].

## Collaboration path and community needs

The challenging program outlined above will require coordinated efforts and collaboration of different theoretical groups working on SRF problems at Old Dominion University, Northwestern University, KEK, Cornell, ANL and others. In addition to the exciting opportunities to establish the fundamental limits of the accelerating gradient and the quality factors of SRF cavities, another outcome of this work would be the development of efficient numerical codes to calculate the field dependence of $Q(H_a)$. This theoretical work will be performed in close collaboration with SRF experimental groups at FNAL, JLab, SLAC, Europe, Japan and China to guide the ongoing optimization of SRF cavities and get feedback from experimental results to help refine the theory and meet the needs of the HEP community. This program can provide an excellent platform for training graduate students and young scientists who will be working at the forefront of theoretical condensed matter research while collaborating with the leading SRF experimental and technological groups at National Laboratories and Universities. Implementation of this research plan will require extensive numerical simulations some of which can hardly be done using the computer workstations at universities and will likely require the access to a much higher level of computational power available in computational centers at National Laboratories.

## Conclusion

Continued investment is required so this fundamental R&D can mature and become project ready for the next generation accelerators.